arXiv:0904.0274v1 [cs.IT] 1 Apr 2009

# Interference Alignment with Asymmetric Complex Signaling - Settling the Høst-Madsen-Nosratinia Conjecture

Viveck Cadambe, Syed A. Jafar, Chenwei Wang*

*E-mail : {vcadambe,syed,chenweiw}@uci.edu*

**Abstract**

It has been conjectured by Høst-Madsen and Nosratinia that complex Gaussian interference channels with constant channel coefficients have only one degree-of-freedom regardless of the number of users. While several examples are known of constant channels that achieve more than 1 degree of freedom, these special cases only span a subset of measure zero. In other words, for almost all channel coefficient values, it is not known if more than 1 degree-of-freedom is achievable. In this paper, we settle the Høst-Madsen-Nosratinia conjecture in the negative. We show that at least 1.2 degrees-of-freedom are achievable for all values of complex channel coefficients except for a subset of measure zero. For the class of linear beamforming and interference alignment schemes considered in this paper, it is also shown that 1.2 is the maximum number of degrees of freedom achievable on the complex Gaussian 3 user interference channel with constant channel coefficients, for almost all values of channel coefficients. To establish the achievability of 1.2 degrees of freedom we introduce the novel idea of asymmetric complex signaling - i.e., the inputs are chosen to be complex but not circularly symmetric. It is shown that unlike Gaussian point-to-point, multiple-access and broadcast channels where circularly *symmetric* complex Gaussian inputs are optimal, for interference channels optimal inputs are in general asymmetric. With asymmetric complex signaling, we also show that the 2 user complex Gaussian $X$ channel with *constant* channel coefficients achieves the outer bound of 4/3 degrees-of-freedom, i.e., the assumption of time-variations/frequency-selectivity used in prior work to establish the same result, is not needed.

*The ordering of authors is alphabetical.

# 1 Introduction

The notion of degrees-of-freedom of a communication network, also known as the capacity pre-log/multiplexing-gain/effective-bandwidth etc., is a fundamental concept in communication and information theory. Intuitively, it measures the number of independent signaling dimensions that are accessible in the network. Degrees-of-freedom characterizations are well known for Gaussian point-to-point, multiple-access, and broadcast channels (with no common messages), with or without multiple-antenna nodes. Much less is known about the degrees of freedom of interference networks, where the distributed nature of the network precludes joint processing of transmitted or received signals. With recent focus on capacity approximations for interference networks, the degrees of freedom characterizations have become especially important. This is because the degrees of freedom characterization also provides a first order capacity approximation, whose accuracy approaches 100% as the signal-to-noise power ratio (SNR) approaches infinity. The high SNR regime is where all desired and interfering signals are much stronger than the local noise at each receiver. The regime is of interest because it directly addresses the problem of interference, believed to be the principal limitation to the capacity of wireless networks.

The study of degrees of freedom of interference networks was pioneered by Høst-Madsen and Nosratinia [1], who showed that the two user interference channel has only one degree of freedom, even with cooperation between transmitters and/or cooperation between receivers, provided this cooperation takes place over Gaussian channels as well. For $K$ user interference channels, Høst-Madsen and Nosratinia showed that it is not possible to achieve more than $K/2$ degrees of freedom. However, it was also conjectured in [1] that the outer bound is loose in general and interference networks have only 1 degree of freedom, regardless of the number of users. Intuitively, the conjecture supports the optimality of orthogonal medium access schemes (e.g. TDMA/FDMA) where each user is assigned a fraction of the channel's degrees of freedom (signaling dimensions) so that the sum of these fractions is equal to one.

Reference [2] showed that the intuition behind the Høst-Madsen-Nosratinia conjecture does not apply to complex (or real) Gaussian $K$ user interference channels with time-varying/frequency-selective channel coefficients. It was shown that for these channels the total number of degrees of freedom is almost surely $K/2$. The key to this surprising result was the new idea of *interference alignment*, introduced for the $X$ channel in [3, 4] and for the interference channel in [2]. In particular, [2] introduced an explicit interference alignment scheme for $K$-user time-varying/frequency-selective interference channel, which comes arbitrarily close to the outer bound of $K/2$ degrees of freedom by coding over sufficiently long symbol extensions. However, since the Høst-Madsen-Nosratinia conjecture was made in the context of interference channels with constant, complex channel coefficients, the conjecture remained open. Evidently, the difficulty lay in determining the feasibility of interference alignment over constant channels.

Interference alignment, as defined in [3], refers to the construction of signals in such a manner that interfering signals cast overlapping shadows at each receiver while the desired signals remain distinct. The key to this construction is the relativity of alignment, i.e. signals align differently at each receiver. Because each receiver sees a different picture, it is possible for the same set of signals to align at one receiver where they are not desired and remain distinct at another receiver where they are desired. The interference alignment schemes proposed in literature can be broadly classified into two categories.

1. Signal Vector Space Alignment Schemes - Linear transmitter precoding and receiver combining operations are used to transform the interference channel into multiple non-interfering

Gaussian channels. The relativity of alignment exploited here is the distinct linear transformation (channel matrix) between each transmitter-receiver pair, which makes sure that signal vectors are rotated differently on each link. The strength of this approach is that these schemes work for all values of channel coefficients with the exception of a subset of measure 0. The limitation here is the need for the assumption that each receiver sees different relative rotations of the input signal vectors. With multiple antennas, the different channel matrices provide these distinct rotations [3, 5]. If multiple antennas are not present, the distinct rotations come from the diagonal channel matrices resulting from multiple channel uses over time-varying/frequency-selective fading channels [2, 6]. However, if the channels are constant (i.e., not time-varying or frequency-selective) then the effective channel matrix resulting from multiple channel uses is simply a scaled identity matrix, which does not rotate the signal vectors at all. Since the signal vectors align the same way at each receiver, interference alignment is not possible without aligning the desired signal with the interference as well. Therefore these schemes have not been effective for interference channels with constant channel coefficients.

2. Signal Level Alignment Schemes - This approach relies on structured coding, e.g., multilevel or lattice codes, to align interference in the signal "level" space. The relativity of alignment exploited here comes from the distinct scaling of signals between each transmitter-receiver pair. Due to the different scaling factors, signal levels align differently at each receiver. Examples of this approach include [7, 8, 9, 10], all of which address constant interference channels. The strength of this approach is its ability to achieve alignment for some constant channels. An apparent weakness may be that since these approaches are derived from the deterministic channel models of [11], they inherit some of the limitations of the deterministic models as well. In particular, deterministic channel models have proven very useful for studying channels with, essentially, real channel coefficients. However, for channels where channel phase and vector alignments play an important part, deterministic models have not been as useful. Another limitation of interference alignment over signal levels is that so far these approaches have been shown to have degrees-of-freedom benefits only for channel coefficients over a subset of measure 0, i.e., only for special cases.

We summarize here the key degrees-of-freedom results for signal level alignment schemes. A multilevel coding based interference alignment scheme was proposed in [9]. The scheme was shown to achieve more than 1 degree of freedom for interference channels where desired channel coefficients are of the form $Q^e$ and interfering channel-coefficients are of the form $Q^o$, where $e, o$ are any even, odd integers, respectively, and $Q$ is a large number (relative to the number of users $K$). However, the special structure assumed on the channel coefficients meant that this scheme was only restricted to channel coefficient values that constitute a subset of measure 0.

Taking the idea of interference alignment in signal-level further, Etkin and Ordentlich [10] proposed a sophisticated lattice alignment scheme for the class of interference channels where the direct channel coefficients are algebraic irrationals and the cross-channel coefficients take rational values. Using results from diophantine approximation theory they showed that a lattice scaled by an algebraic irrational factor "stands out" from a lattice scaled by a rational factor allowing a separation of signal and interference. This scheme proved the achievability of the full $K/2$ degrees of freedom for a dense set of channel coefficients. However, the assumptions on the channel coefficient values (e.g. rationals and algebraic irrationals) restricted its scope to, once again, a subset of measure zero, and the validity of Høst-Madsen-Nosratinia conjecture remained unknown for almost all channel coefficient values.

While much of the interference alignment work for constant channels has focused on achieving more than 1 degree-of-freedom, i.e., there are at least two results that provide a counterpoint by identifying conditions under which the degrees of freedom of a $K$ user interference channel may be limited to values less than $K/2$. Reference [12] provides conditions under which a fully connected 3 user complex Gaussian interference channel with constant coefficients can have only 1 degree of freedom. The conditions are re-stated in this paper in Theorem 3 to provide a comparison with analogous conditions that emerge out of this work. Another limiting result, shown in [10], is that for real Gaussian interference channels where all channel coefficients are rational, $K/2$ degrees of freedom are not achievable. Like the alignment schemes, these results are also limited to channel coefficient values over a subset of measure 0.

With one exception, the importance of channel phase is ignored in nearly all of the interference alignment schemes presented so far. The exception comes from [2] where the following example is presented to introduce the concept of interference alignment.

*Phase Alignment Example [2]:* Consider a symmetric interference channel where all direct channel-coefficients are equal to 1 and all cross-channel coefficients are equal to $j = \sqrt{-1}$. If the additive white Gaussian noise power at each receiver is normalized to unity and the transmitted signal power at each transmitter is limited to SNR, then the exact sum-capacity of this interference channel is shown to be $\frac{K}{2}\log(1 + 2\text{SNR})$ bits/channel-use.

The phase-alignment example described above, is the starting point for a new direction that we pursue in this paper. First, we note that wireless channels are invariably modeled with complex (instead of real) channel coefficients, inputs and noise to capture both the in-phase and quadrature-phase signaling dimensions. Moreover, the Høst-Madsen-Nosratinia conjecture is made in the context of complex channels and therefore must be proved or disproved in the same setting. Further, the complex model offers a richer signal space and therefore may not suffer from some of the degrees-of-freedom limitations associated with real models. For example, while [10] shows that $K/2$ degrees of freedom are not achievable with rational coefficients in the real Gaussian interference channel, the phase-alignment example shows that the complex Gaussian interference channel, even with coefficients that have rational (in fact, integer) real and imaginary parts, can still achieve the full $K/2$ degrees of freedom. Unfortunately, like all other examples described earlier, the phase alignment example of [2] is also restricted to a special choice of channel coefficients and leaves the Høst-Madsen-Nosratinia conjecture open for almost all channel coefficients.

## 1.1 A New Idea - Asymmetric Complex Signaling

In this section we summarize an important new idea that emerges from this work - the need for asymmetric complex signaling.

Consider the input optimization problem for interference networks. Suppose we restrict the achievable schemes to Gaussian inputs. Because we do not have multiple antennas, there is no input covariance matrix to optimize. The input optimization therefore appears to be limited to only a power optimization. Now consider the following two questions.

- **Symbol Extensions** - *Can we do better by transforming the complex scalar input optimization problem to a complex vector input optimization problem by considering multiple, say $M$, channel uses as one $M$-dimensional complex super-symbol?*

  Input optimization in this case becomes the problem of optimizing the $M \times M$ input covariance matrix of the $M$ dimensional complex input Gaussian vector.

- **Asymmetric Complex Signaling** - *Can we do better by transforming the $M$ dimensional complex system to a $2M$ dimensional real system and optimizing inputs over the $2M$ real dimensions?*

  Input optimization in this case becomes the problem of optimizing the $2M \times 2M$ input covariance matrix of the $2M$ dimensional real input Gaussian vector.

Since the channels are constant across channel uses, the intuitive answer here may be that symbol extensions are not going to be useful. Indeed we do not see any benefits of symbol extensions in point to point, MAC or BC channels, even in the MIMO setting. Interestingly symbol extensions do help, even with constant channel coefficients, in the MIMO compound broadcast channel [13], the MIMO X channel [3] and the MIMO interference channel [2]. However, in our case since we do not have multiple antennas it is not immediately obvious if symbol extensions will be useful.

The second possibility, of asymmetric complex signaling, goes against the generic intuition that favors circularly symmetric Gaussians. In wireless communication theory we typically come across only circularly *symmetric* complex Gaussian random variables. The additive noise is invariably modeled as circularly symmetric complex Gaussian. The most commonly studied channel fading model, Rayleigh fading, refers to circularly symmetric complex Gaussian channel coefficient values. More importantly, since our interest is in optimal (capacity-achieving) input distributions, circularly symmetric complex Gaussians are omnipresent there as well. For complex Gaussian point-to-point, multiple-access and broadcast channels with constant channel coefficients, with or without multiple antennas, capacity-achieving input distributions are circularly symmetric complex Gaussian. Intuitively, the reason is that circularly symmetric complex Gaussian random variables maximize entropy for a given second moment [14]. We are not aware of any works on capacity/rate optimization for complex Gaussian wireless networks where asymmetric Gaussian inputs outperform circularly symmetric Gaussian inputs — with one important exception, and that brings us back to the phase alignment example of [2].

The capacity achieving scheme for the phase-alignment example requires each transmitter to use only *real* valued Gaussian inputs, as opposed to circularly symmetric complex Gaussian inputs. This choice of input signals ensures that interference at each receiver aligns in the imaginary dimension while the desired signal is received free from interference in the real dimension of the complex received signal space. However, since the phase-alignment example assumes very specific values of channel coefficients, it is also not obvious if it extends to arbitrary values of channel coefficients.

Aside from settling the Høst-Madsen-Nosratinia conjecture, the main contribution of this work is to establish the need for asymmetric complex signaling (and symbol extensions), not only for some special cases, but for almost all values of complex channel coefficients. The achievable scheme proposed in paper relies on both channel extensions and asymmetric complex signaling, and is shown to achieve at least 1.2 degrees of freedom for almost all complex Gaussian interference channels with 3 or more users. Notably, circularly symmetric Gaussian inputs can only achieve 1 degree of freedom on this channel. Further, because our achievable scheme uses only simple beamforming with every receiver treating interference as noise, it shows that asymmetric complex signaling and symbol extensions are important not only for capacity characterizations but also for practically motivated rate optimization problems where the receivers do not have multi-user detection capabilities, e.g. [15].

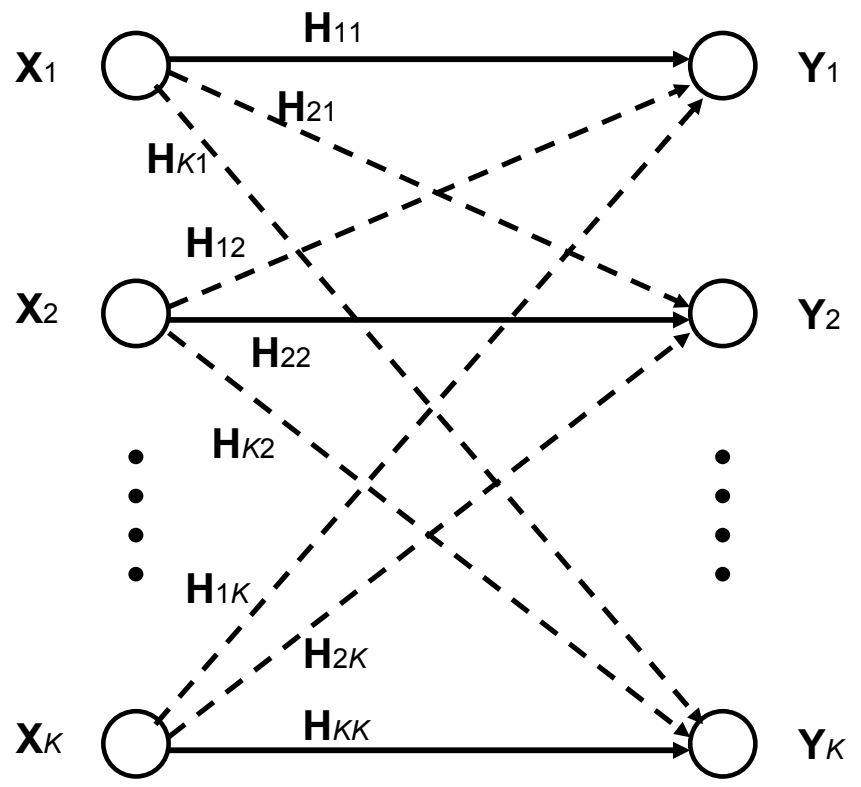

Figure 1: $K$ User Interference Channel

### 1.2 Channel Model

For the complex Gaussian interference channel with $K$ users, the signal received at receiver $r$ during the $n^{th}$ channel use, is expressed as

$$\mathbf{Y}_r(n) \;=\; \sum_{t=1}^{K} \mathbf{H}_{rt}\mathbf{X}_t(n) + \mathbf{Z}_r(n) \tag{1}$$

$\mathbf{Z}_r(n)$ represents independent identically distributed (i.i.d.) zero mean unit variance circularly symmetric complex Gaussian noise terms. $\mathbf{X}_t(n)$ is the signal sent from transmitter $t$. $\mathbf{H}_{rt} = |\mathbf{H}_{rt}|e^{j\phi_{rt}}$ is the complex channel coefficient between transmitter $t$ and receiver $r$, whose value is held constant across channel uses. All nodes have only a single antenna each, so that the signals, channel coefficients and noise are complex *scalars*. The transmit power constraint is represented as:

$$\mathrm{E}\left[|\mathbf{X}_t|^2\right] \leq \mathrm{SNR}, \quad t \in \{1, 2, \cdots, K\}. \tag{2}$$

As usual, in the $K$ user interference channel, transmitter $k$ has message $W_k$ for receiver $k$, $k \in \{1, 2, \cdots, K\}$. All messages are independent. The probability of error $P_e$, achievable rates $R_1, R_2, \cdots, R_K$ and sum-capacity $C_\Sigma(\mathrm{SNR})$ of the interference channel are defined in the standard Shannon sense. The number of degrees of freedom $d$ is defined as:

$$d \;=\; \lim_{\mathrm{SNR}\to\infty} \frac{C_\Sigma(\mathrm{SNR})}{\log(\mathrm{SNR})} \tag{3}$$

We also use an alternative representation for equation (1) in terms of only real quantities as:

$$\underbrace{\begin{bmatrix} \mathrm{Re}\{\mathbf{Y}_r(n)\} \\ \mathrm{Im}\{\mathbf{Y}_r(n)\} \end{bmatrix}}_{Y_r(n)} \;=\; \sum_{t=1}^{K} \underbrace{\overbrace{|\mathbf{H}_{rt}|}^{h_{rt}} \overbrace{\begin{bmatrix} \cos(\phi_{rt}) & -\sin(\phi_{rt}) \\ \sin(\phi_{rt}) & \cos(\phi_{rt}) \end{bmatrix}}^{U(\phi_{rt})}}_{H_{rt}} \underbrace{\begin{bmatrix} \mathrm{Re}\{\mathbf{X}_t(n)\} \\ \mathrm{Im}\{\mathbf{X}_t(n)\} \end{bmatrix}}_{X_t(n)} + \underbrace{\begin{bmatrix} \mathrm{Re}\{\mathbf{Z}_r(n)\} \\ \mathrm{Im}\{\mathbf{Z}_r(n)\} \end{bmatrix}}_{Z_r(n)} \tag{4}$$

$$\Rightarrow Y_r(n) \;=\; \sum_{t=1}^{K} H_{rt}X_t(n) + Z_r(n) = \sum_{t=1}^{K} h_{rt}U(\phi_{rt})X_t(n) + Z_r(n) \tag{5}$$

Thus, bold font is reserved for complex quantities while the real representations of the same variables use normal font. Note that while the complex quantities, e.g., $\mathbf{Y}_r(n), \mathbf{H}_{rt}$, are scalars, the real counterparts, $Y_r(n), H_{rt}$, etc., are matrices. $U(\phi)$ is a rotation matrix with the properties:

$$U(\phi)^{-1} = U(-\phi)$$
$$U(\theta)U(\phi) = U(\phi)U(\theta) = U(\phi + \theta)$$

To avoid cumbersome notation, we will drop the channel-use index "$n$" unless necessary to avoid ambiguity.

With the exception of Corollary 1 (which is a trivial generalization of Theorem 2 to more than 3 users) in this paper we focus primarily on the $K = 3$ user interference channel.

## 2 Phase Alignment

For the 3 user constant MIMO interference channel where each node is equipped with $M = 2$ antennas, an explicit interference alignment solution is found in [2] that achieves the outer bound of 3/2 degrees of freedom. For our case, we have only single antenna nodes. However, viewing complex numbers as two-dimensional vectors, the channel input-output equations (4) are analogous to an interference channel where each node is equipped with two antennas. A natural question then is to determine if the MIMO interference alignment solution can directly translate into a generalized phase alignment scheme for all channel coefficients over a subset of non-zero measure. In this section, we answer this question in the negative.

In order to achieve a total of 3/2 degrees of freedom, each user in the 3 user interference channel must achieve 1 degree of freedom over a 2 dimensional space. Since the equations (4) already represent a 2 dimensional space, we only need to design the signal vectors along which each user can achieve 1 degree of freedom. In other words, we need to design the *real* vectors $V_1, V_2, V_3$, each of dimension $2 \times 1$ such that:

$$\begin{aligned} X_1 &= V_1 x_1 \\ X_2 &= V_2 x_2 \\ X_3 &= V_3 x_3 \end{aligned}$$

Here, $V_1, V_2, V_3$ are the precoding vectors, optimized for the channel coefficient values, but independent of the messages, while $x_1, x_2, x_3$ represent the real-scalar codewords which carry the messages.

Now consider receiver 1. The desired signal is received along the vector $H_{11}V_1$ while interference arrives along the vectors $H_{12}V_2$ and $H_{13}V_3$. In a 2 dimensional signal space, in order to leave one interference-free dimension for the desired signal, the two interfering signals must span only a one-dimensional space. This means:

$$\begin{aligned} \text{span}(H_{12}V_2) &= \text{span}(H_{13}V_3) && (6) \\ \Rightarrow \text{span}(h_{12}U(\phi_{12})V_2) &= \text{span}(h_{13}U(\phi_{13})V_3) && (7) \\ \Rightarrow \text{span}(U(\phi_{12})V_2) &= \text{span}(U(\phi_{13})V_3) && (8) \\ \Rightarrow \text{span}(V_2) &= \text{span}(U(\phi_{12})^{-1}U(\phi_{13})V_3) && (9) \\ \Rightarrow \text{span}(V_2) &= \text{span}(U(\phi_{13} - \phi_{12})V_3) && (10) \end{aligned}$$

Similarly, at receiver 2, interference from transmitters 1 and 3 must align,

$$\begin{aligned} \text{span}(H_{21}V_1) &= \text{span}(H_{23}V_3) && (11) \\ \Rightarrow \text{span}(V_3) &= \text{span}(U(\phi_{21}-\phi_{23})V_1) && (12) \end{aligned}$$

and at receiver 3, interference from transmitters 1 and 2 must align,

$$\begin{aligned} \text{span}(H_{31}V_1) &= \text{span}(H_{32}V_2) && (13) \\ \Rightarrow \text{span}(V_1) &= \text{span}(U(\phi_{32}-\phi_{31})V_2) && (14) \\ \Rightarrow \text{span}(V_1) &= \text{span}(U(\phi_{32}-\phi_{31}+\phi_{13}-\phi_{12})V_3) && (15) \\ \Rightarrow \text{span}(V_1) &= \text{span}(U(\phi_{32}-\phi_{31}+\phi_{13}-\phi_{12}+\phi_{21}-\phi_{23})V_1) && (16) \\ \Rightarrow V_1 &= \text{eigenvec}(U(\phi_{32}-\phi_{31}+\phi_{13}-\phi_{12}+\phi_{21}-\phi_{23})) && (17) \end{aligned}$$

(15) is obtained by substituting (10) into (14). (16) is obtained by substituting (12) into (15). The solution is formalized in the following theorem.

**Theorem 1** *The 3-user complex Gaussian interference channel with constant channel coefficients has* 3/2 *degrees of freedom if*

$$\begin{aligned} \phi_{32}+\phi_{21}+\phi_{13}-(\phi_{12}+\phi_{23}+\phi_{31}) &= 0 \ mod(\pi) && (18) \\ \phi_{21}-\phi_{23}+\phi_{13}-\phi_{11} &\neq 0 \ mod(\pi) && (19) \\ \phi_{22}+\phi_{13}-\phi_{12}-\phi_{23} &\neq 0 \ mod(\pi) && (20) \\ \phi_{33}+\phi_{21}-\phi_{23}-\phi_{31} &\neq 0 \ mod(\pi) && (21) \end{aligned}$$

*where $a = 0$ $mod(\pi)$ if and only if $a$ is an integer multiple of $\pi$.*

*Proof:* Based on our channel model (4) the vector $V_1$ must have real elements. (16) requires that the real vector $V_1$ is an eigenvector of a rotation matrix $U(\phi)$. The rotation matrix $U(\phi)$ has real eigenvectors only if $\phi$ is an integer multiple of $\pi$. This gives us condition (18). The remaining conditions are easily verified to be necessary to make sure that the desired signal vector is linearly independent of the interference vector at each receiver. ■

*Remark:* Because of the constraint (18), the solution is once again restricted to a subset of channel coefficient values that has measure 0 and the validity of the Høst-Madsen-Nosratinia conjecture is not determined.

## 3 Achievability of 1.2 Degrees of Freedom

Consider the 5 symbol extension of the 3 user complex Gaussian interference channel with constant channel coefficients.

$$\underbrace{\begin{bmatrix} \mathbf{Y}_r(5\overline{n}+1) \\ \mathbf{Y}_r(5\overline{n}+2) \\ \vdots \\ \mathbf{Y}_r(5(\overline{n}+1)) \end{bmatrix}}_{\overline{\mathbf{Y}}_r(\overline{n}):5\times 1} = \sum_{t=1}^{3} \mathbf{H}_{rt} \underbrace{\begin{bmatrix} \mathbf{X}_t(5\overline{n}+1) \\ \mathbf{X}_t(5\overline{n}+2) \\ \vdots \\ \mathbf{X}_t(5(\overline{n}+1)) \end{bmatrix}}_{\overline{\mathbf{X}}_t(\overline{n}):5\times 1} + \underbrace{\begin{bmatrix} \mathbf{Z}_r(5\overline{n}+1) \\ \mathbf{Z}_r(5\overline{n}+2) \\ \vdots \\ \mathbf{Z}_r(5(\overline{n}+1)) \end{bmatrix}}_{\overline{\mathbf{Z}}_r(\overline{n}):5\times 1} \qquad (22)$$

Thus, we have a 5 dimensional complex signal space, or equivalently, a 10 dimensional real signal space.

$$\underbrace{\begin{bmatrix} \text{Re}\{\mathbf{Y}_r(5\overline{n}+1)\} \\ \text{Im}\{\mathbf{Y}_r(5\overline{n}+1)\} \\ \text{Re}\{\mathbf{Y}_r(5\overline{n}+2)\} \\ \vdots \\ \text{Im}\{\mathbf{Y}_r(5(\overline{n}+1))\} \end{bmatrix}}_{\overline{Y}_r(\overline{n}):10\times 1} = \sum_{t=1}^{3} h_{rt} \underbrace{(U(\phi_{rt}) \otimes I_{5\times 5})}_{\overline{U}(\phi_{rt}):10\times 10} \underbrace{\begin{bmatrix} \text{Re}\{\mathbf{X}_t(5\overline{n}+1)\} \\ \text{Im}\{\mathbf{X}_t(5\overline{n}+1)\} \\ \text{Re}\{\mathbf{X}_t(5\overline{n}+2)\} \\ \vdots \\ \text{Im}\{\mathbf{X}_t(5(\overline{n}+1))\} \end{bmatrix}}_{\overline{X}_t(\overline{n}):10\times 1} + \underbrace{\begin{bmatrix} \text{Re}\{\mathbf{Z}_r(5\overline{n}+1)\} \\ \text{Im}\{\mathbf{Z}_r(5\overline{n}+1)\} \\ \text{Re}\{\mathbf{Z}_r(5\overline{n}+2)\} \\ \vdots \\ \text{Im}\{\mathbf{Z}_r(5(\overline{n}+1))\} \end{bmatrix}}_{\overline{Z}_r(\overline{n}):10\times 1}$$

where $\otimes$ indicates the Kronecker product operation. $\overline{U}(\phi)$ is a block-diagonal matrix with the $2\times 2$ block $U(\phi)$ repeated along the main diagonal. Clearly, $\overline{U}(\phi)$ also satisfies the properties:

$$\overline{U}(\phi)^{-1} = \overline{U}(-\phi) \tag{23}$$

$$\overline{U}(\theta)U(\phi) = \overline{U}(\phi)\overline{U}(\theta) = U(\phi+\theta) \tag{24}$$

Within this 10 dimensional real signal space, each transmitter sends 4 separately encoded real streams along 4 linearly independent real vectors.

$$\overline{X}_t(\overline{n}) \;=\; \sum_{s=1}^{4} \overline{V}_t^s x_t^s(\overline{n})$$

For example, $\overline{V}_1^1, \overline{V}_1^2, \overline{V}_1^3, \overline{V}_1^4$ are the four signaling vectors used by transmitter 1 to send 4 separately encoded scalar real codeword symbols $x_1^1(\overline{n}), x_1^2(\overline{n}), x_1^3(\overline{n}), x_1^4(\overline{n})$. The signal vectors for each transmitter are similarly defined. Since each transmitter sends 4 streams, the total number of streams sent is 12. Sending 12 real streams over 10 real symbols, or equivalently 6 complex streams over 5 complex symbols, means that if these streams can be separated from the interference and from each other, then a total of $6/5 = 1.2$ degrees of freedom are achieved per channel use.

Interference alignment is needed to accomplish this objective. Consider receiver 1. Out of the 10 real dimensions available to the receiver, 4 are needed for his desired signals, leaving no more than 6 dimensions for interference. Since there are 8 interfering signals we need two alignments at each receiver to make sure that interference spans only 6 real dimensions. For receiver 1 we choose the following.

$$\overline{U}(\phi_{12})\overline{V}_2^3 \;=\; \overline{U}(\phi_{13})\overline{V}_3^1 \tag{25}$$

$$\overline{U}(\phi_{13})\overline{V}_3^4 \;=\; \overline{U}(\phi_{12})\overline{V}_2^2 \tag{26}$$

For receiver 2 we choose the following alignments.

$$\overline{U}(\phi_{23})\overline{V}_3^3 \;=\; \overline{U}(\phi_{21})\overline{V}_1^1 \tag{27}$$

$$\overline{U}(\phi_{21})\overline{V}_1^4 \;=\; \overline{U}(\phi_{23})\overline{V}_3^2 \tag{28}$$

Similarly, for receiver 3 we choose the following alignments.

$$\overline{U}(\phi_{31})\overline{V}_1^3 \;=\; \overline{U}(\phi_{32})\overline{V}_2^1 \tag{29}$$

$$\overline{U}(\phi_{32})\overline{V}_2^4 \;=\; \overline{U}(\phi_{31})\overline{V}_1^2 \tag{30}$$

Equations (25)-(30) ensure that at each receiver, interference cannot span more than 6 dimensions.

Suppose at each transmitter $t = 1, 2, 3$, we choose the first two $10 \times 1$ signaling vectors $\overline{V}_t^1$, and $\overline{V}_t^2$ randomly according to some continuous distribution. The remaining two signaling vectors at each transmitter $\overline{V}_t^3, \overline{V}_t^4$ are chosen according to (25)-(30). This ensures interference alignment.

What remains to be shown is that at each receiver the desired signal vectors are linearly independent among themselves and also from the interference. Without loss of generality we show this for receiver 1. The same argument applies at each receiver due to the symmetry of the signaling scheme.

$$\overline{Y}_1 = \sum_{s=1}^{4} (h_{11}x_1^s)\,\overline{U}(\phi_{11})\overline{V}_1^s + \sum_{s=1}^{4} (h_{12}x_2^s)\,\overline{U}(\phi_{12})\overline{V}_2^s + \sum_{s=1}^{4} (h_{13}x_3^s)\,\overline{U}(\phi_{13})\overline{V}_3^s \quad (31)$$

$$\begin{aligned}\overline{Y}_1 &= \sum_{s=1}^{4} (h_{11}x_1^s)\,\overline{U}(\phi_{11})\overline{V}_1^s + \sum_{s\in\{1,4\}} (h_{12}x_2^s)\,\overline{U}(\phi_{12})\overline{V}_2^s + \sum_{s\in\{2,3\}} (h_{13}x_3^s)\,\overline{U}(\phi_{13})\overline{V}_3^s \\ &\quad +(h_{12}x_2^2 + h_{13}x_3^4)\overline{U}(\phi_{12})\overline{V}_2^2 + (h_{13}x_3^1 + h_{23}x_2^3)\overline{U}(\phi_{13})\overline{V}_3^1 \quad (32)\end{aligned}$$

Thus, the desired signals arrive along the following 4, real, $10 \times 1$ vectors.

$$\overline{U}(\phi_{11})\overline{V}_1^1,\ \overline{U}(\phi_{11})\overline{V}_1^2,\ \overline{U}(\phi_{11})\overline{V}_1^3,\ \overline{U}(\phi_{11})\overline{V}_1^4 \quad (33)$$

The 8 interfering signals arrive along the following 6, real, $10 \times 1$ vectors.

$$\overline{U}(\phi_{12})\overline{V}_2^1,\ \overline{U}(\phi_{12})\overline{V}_2^2,\ \overline{U}(\phi_{12})\overline{V}_2^4,\ \overline{U}(\phi_{13})\overline{V}_3^1,\ \overline{U}(\phi_{13})\overline{V}_3^2,\ \overline{U}(\phi_{13})\overline{V}_3^3 \quad (34)$$

In order to show that the desired signal vectors are linearly independent among themselves and also from the interference, it suffices to show that all 10 vectors (4 desired and 6 interference carrying vectors) are linearly independent. Suppose, there exist 10 real constants $a_1, a_2, \cdots, a_{10}$ such that

$$\begin{aligned}0 &= a_1\overline{U}(\phi_{11})\overline{V}_1^1 + a_2\overline{U}(\phi_{11})\overline{V}_1^2 + a_3\overline{U}(\phi_{11})\overline{V}_1^3 + a_4\overline{U}(\phi_{11})\overline{V}_1^4 + a_5\overline{U}(\phi_{12})\overline{V}_2^1 + a_6\overline{U}(\phi_{12})\overline{V}_2^2 \\ &\quad +a_7\overline{U}(\phi_{12})\overline{V}_2^4 + a_8\overline{U}(\phi_{13})\overline{V}_3^1 + a_9\overline{U}(\phi_{13})\overline{V}_3^2 + a_{10}\overline{U}(\phi_{13})\overline{V}_3^3 \quad (35)\end{aligned}$$

then we show that all 10 constants must be zero

$$a_i = 0, \forall i \in \{1, 2, \cdots, 10\} \quad (36)$$

which establishes the linear independence of the 10 vectors. Using (25)-(30), we can rewrite (35) in terms of only vectors $\overline{V}_t^1, \overline{V}_t^2$, as

$$\begin{aligned}0 &= a_1\overline{U}(\phi_{11})\overline{V}_1^1 + a_2\overline{U}(\phi_{11})\overline{V}_1^2 + a_3\overline{U}(\phi_{11} + \phi_{32} - \phi_{31})\overline{V}_2^1 + a_4\overline{U}(\phi_{11} + \phi_{23} - \phi_{21})\overline{V}_3^2 \\ &\quad +a_5\overline{U}(\phi_{12})\overline{V}_2^1 + a_6\overline{U}(\phi_{12})\overline{V}_2^2 + a_7\overline{U}(\phi_{12} + \phi_{31} - \phi_{32})\overline{V}_1^2 + a_8\overline{U}(\phi_{13})\overline{V}_3^1 + a_9\overline{U}(\phi_{13})\overline{V}_3^2 \\ &\quad +a_{10}\overline{U}(\phi_{13} + \phi_{21} - \phi_{23})\overline{V}_1^1 \quad (37) \\ &= \left[a_1 I + a_{10}\overline{U}(\phi_{13} + \phi_{21} - \phi_{23} - \phi_{11})\right]\overline{U}(\phi_{11})\overline{V}_1^1 + \left[a_2 I + a_7\overline{U}(\phi_{12} + \phi_{31} - \phi_{32} - \phi_{11})\right]\overline{U}(\phi_{11})\overline{V}_1^2 \\ &\quad + \left[a_5 I + a_3\overline{U}(\phi_{11} + \phi_{32} - \phi_{31} - \phi_{12})\right]\overline{U}(\phi_{12})\overline{V}_2^1 + a_6\overline{U}(\phi_{12})\overline{V}_2^2 \\ &\quad +a_8\overline{U}(\phi_{13})\overline{V}_3^1 + \left[a_9 I + a_4\overline{U}(\phi_{11} + \phi_{23} - \phi_{21} - \phi_{13})\right]\overline{U}(\phi_{13})\overline{V}_3^2 \quad (38)\end{aligned}$$

Since the vectors $\overline{V}_2^2, \overline{V}_3^1$ are generated independently of the remaining terms, they are in general position, i.e. linearly independent of the remaining terms with probability 1. This implies that

$$\begin{aligned} a_6 &= 0 && (39)\\ a_8 &= 0 && (40) \end{aligned}$$

With the remaining terms, equation (38) is equivalently represented as

$$\begin{aligned} 0 \;=\; & (a_1 + a_{10}e^{j(\phi_{13}+\phi_{21}-\phi_{23}-\phi_{11})})e^{j\phi_{11}}\overline{\mathbf{V}}_1^1 + (a_2 + a_7 e^{j(\phi_{12}+\phi_{31}-\phi_{32}-\phi_{11})})e^{j\phi_{11}}\overline{\mathbf{V}}_1^2 \\ & + (a_5 + a_3 e^{j(\phi_{11}+\phi_{32}-\phi_{31}-\phi_{12})})e^{j\phi_{12}}\overline{\mathbf{V}}_2^1 + (a_9 + a_4 e^{j(\phi_{11}+\phi_{23}-\phi_{21}-\phi_{13})})e^{j\phi_{13}}\overline{\mathbf{V}}_3^2 \qquad (41) \end{aligned}$$

where

$$\overline{V}_t^s = \begin{bmatrix} \mathrm{Re}\{\overline{\mathbf{V}}_t^s[1]\} \\ \mathrm{Im}\{\overline{\mathbf{V}}_t^s[1]\} \\ \mathrm{Re}\{\overline{\mathbf{V}}_t^s[2]\} \\ \mathrm{Im}\{\overline{\mathbf{V}}_t^s[2]\} \\ \vdots \\ \mathrm{Re}\{\overline{\mathbf{V}}_t^s[5]\} \\ \mathrm{Im}\{\overline{\mathbf{V}}_t^s[5]\} \end{bmatrix}, \quad t \in \{1,2,3\}, s \in \{1,2,3,4\} \qquad (42)$$

Since the $5 \times 1$ complex vectors $\overline{\mathbf{V}}_1^1, \overline{\mathbf{V}}_1^2, \overline{\mathbf{V}}_2^1, \overline{\mathbf{V}}_3^2$ are all independently generated and there are only 4 of them, they are linearly independent with probability 1. Thus, we must have:

$$\begin{aligned} 0 &= a_1 + a_{10}e^{j(\phi_{13}+\phi_{21}-\phi_{23}-\phi_{11})} && (43)\\ 0 &= a_2 + a_7 e^{j(\phi_{12}+\phi_{31}-\phi_{32}-\phi_{11})} && (44)\\ 0 &= a_5 + a_3 e^{j(\phi_{11}+\phi_{32}-\phi_{31}-\phi_{12})} && (45)\\ 0 &= a_9 + a_4 e^{j(\phi_{11}+\phi_{23}-\phi_{21}-\phi_{13})} && (46) \end{aligned}$$

Since the $a_i$ are all real, equating the imaginary parts of equations (43)-(46) results in the following.

$$\begin{aligned} 0 &= a_{10}\sin(\phi_{13}+\phi_{21}-\phi_{23}-\phi_{11}) && (47)\\ 0 &= a_7\sin(\phi_{12}+\phi_{31}-\phi_{32}-\phi_{11}) && (48)\\ 0 &= a_3\sin(\phi_{11}+\phi_{32}-\phi_{31}-\phi_{12}) && (49)\\ 0 &= a_4\sin(\phi_{11}+\phi_{23}-\phi_{21}-\phi_{13}) && (50) \end{aligned}$$

Suppose the arguments of the $\sin(\cdot)$ functions are not-integer multiples of $\pi$. Then all $a_i$ must be zero. This proves that the *real* signal vectors at receiver 1 are linearly independent among themselves and from the interference-subspace. Thus, each desired signal vector can be projected into the null space of the rest of the desired and interfering signal vectors to achieve one-degree of freedom per desired signal vector. By symmetry, the same arguments can be applied to each receiver. Overall, we are able to achieve 12/10 degrees of freedom. Thus we have shown the main result of this paper, as stated in the following theorem.

**Theorem 2** *The* 3 *user complex Gaussian interference channel with constant coefficients, defined in Section 1.2, achieves* 1.2 *degrees of freedom if all of the following conditions are satisfied*

$$\phi_{13} + \phi_{21} - \phi_{23} - \phi_{11} \neq 0 \ mod \ (\pi) \tag{51}$$
$$\phi_{12} + \phi_{31} - \phi_{32} - \phi_{11} \neq 0 \ mod \ (\pi) \tag{52}$$
$$\phi_{21} + \phi_{32} - \phi_{31} - \phi_{22} \neq 0 \ mod \ (\pi) \tag{53}$$
$$\phi_{23} + \phi_{12} - \phi_{13} - \phi_{22} \neq 0 \ mod \ (\pi) \tag{54}$$
$$\phi_{32} + \phi_{13} - \phi_{12} - \phi_{33} \neq 0 \ mod \ (\pi) \tag{55}$$
$$\phi_{31} + \phi_{23} - \phi_{21} - \phi_{33} \neq 0 \ mod \ (\pi) \tag{56}$$

**Corollary 1** *The $K$ user complex Gaussian interference channel with constant coefficients, defined in Section 1.2, has at least* 1.2 *degrees of freedom for almost all values of channel coefficients.*

*Proof:* Since the exceptions (51)-(56) represent a subset of channel coefficients of measure 0, Theorem 2 implies Corollary 1 in the 3 user case. The generalization to the $K$ user interference channel is trivial because 1.2 degrees of freedom are achievable by simply shutting off all but 3 users. ■

Thus, Corollary 1 settles the Høst-Madsen-Nosratinia conjecture in the negative. The exception-conditions (51)-(56) have interesting similarties to the following singularity conditions identified in [12] and re-stated here in our context.

**Theorem 3** *[12] The* 3 *user complex Gaussian interference channel with constant coefficients, defined in Section 1.2, has only* 1 *degree of freedom if any of the following conditions is satisfied.*

$$\textit{Condition 1:} \ \frac{h_{13}h_{21}}{h_{23}h_{11}} = 1 \quad \textit{and} \quad \phi_{13} + \phi_{21} - \phi_{23} - \phi_{11} = 0 \ mod \ (2\pi) \tag{57}$$
$$\textit{Condition 2:} \ \frac{h_{12}h_{31}}{h_{32}h_{11}} = 1 \quad \textit{and} \quad \phi_{12} + \phi_{31} - \phi_{32} - \phi_{11} = 0 \ mod \ (2\pi) \tag{58}$$
$$\textit{Condition 3:} \ \frac{h_{21}h_{32}}{h_{31}h_{22}} = 1 \quad \textit{and} \quad \phi_{21} + \phi_{32} - \phi_{31} - \phi_{22} = 0 \ mod \ (2\pi) \tag{59}$$
$$\textit{Condition 4:} \ \frac{h_{23}h_{12}}{h_{13}h_{22}} = 1 \quad \textit{and} \quad \phi_{23} + \phi_{12} - \phi_{13} - \phi_{22} = 0 \ mod \ (2\pi) \tag{60}$$
$$\textit{Condition 5:} \ \frac{h_{32}h_{13}}{h_{12}h_{33}} = 1 \quad \textit{and} \quad \phi_{32} + \phi_{13} - \phi_{12} - \phi_{33} = 0 \ mod \ (2\pi) \tag{61}$$
$$\textit{Condition 6:} \ \frac{h_{31}h_{23}}{h_{21}h_{33}} = 1 \quad \textit{and} \quad \phi_{31} + \phi_{23} - \phi_{21} - \phi_{33} = 0 \ mod \ (2\pi) \tag{62}$$

It is interesting to note that the same phase expressions appear in theorems 2 and 3. Consider, for example, the interference channel where all channel coefficients have magnitude 1, i.e. $h_{rt} = 1, \forall r, t \in \{1, 2, 3\}$. Then, theorems 2 and 3 can be used to identify all channels (i.e. the channels that have only 1 degree of freedom), except those cases where at least one of the phase expressions is an odd multiple of $\pi$ and none of the phase expressions is an even multiple of $\pi$. One such scenario is the 3 user interference channel with $\mathbf{H}_{rt} = 1$ if $r = t$ and $\mathbf{H}_{rt} = -1$ if $r \neq t$, $\forall r, t \in \{1, 2, 3\}$.

# 4 Upper bound

The best known degrees-of-freedom outer bound for the fully connected 3-user complex Gaussian interference channel with constant coefficients is $\frac{3}{2}$. Stronger outer bounds are only known for special cases, such as Theorem 3 and reference [10] where it is shown that 3/2 degrees of freedom are not achievable when all channel coefficients are real and rational.

In this section we show that the class of linear interference alignment schemes considered in this work cannot achieve more than 1.2 degrees of freedom for almost all complex Gaussian 3-user interference channels with constant channel coefficients. Note that this does not preclude the existence of other schemes that may surpass 1.2 degrees of freedom, and even achieve the outer bound of 3/2 degrees of freedom. In fact the existence of such schemes is already shown in [9, 10] as well as in Theorem 1 in this paper. However, all these cases constitute a subset of measure 0 over the set of all possible values of complex channel coefficients.

**Lemma 1** *For any given complex vector* $\mathbf{V}$*, and for any given angles* $\phi, \beta$*, such that,*

$$\sin(\alpha - \beta) \neq 0, \tag{63}$$

*there exist real constants* $(c_1, c_2) \in \mathbb{R}^2$ *such that*

$$\mathbf{V} = c_1 \mathbf{V} e^{j\alpha} + c_2 \mathbf{V} e^{j\beta} \tag{64}$$

*Proof:* It suffices to show that there exist real constants $(c_1, c_2)$ such that

$$1 - c_1 e^{j\alpha} - c_2 e^{j\beta} = 0 \tag{65}$$

Writing the real and imaginary parts separately, we have the following equations.

$$c_1 \cos(\alpha) + c_2 \cos(\beta) = 1 \tag{66}$$
$$c_1 \sin(\alpha) + c_2 \sin(\beta) = 0 \tag{67}$$

We have two real linear equations in two real variables $c_1, c_2$. A solution exists if the matrix

$$\begin{bmatrix} \cos(\alpha) & \cos(\beta) \\ \sin(\alpha) & \sin(\beta) \end{bmatrix} \tag{68}$$

is invertible, i.e., has a non-zero determinant. But the determinant of this matrix is $\sin(\beta - \alpha)$, which is guaranteed to be non-zero by (63). ■

### 4.1 Limitations of the Linear Interference Alignment Scheme

Consider a generalization of the interference alignment scheme used in Section 3. Instead of a 5 symbol extension, suppose we take an $S$ symbol extension, so that the total number of signaling dimensions available at each transmitter or receiver is equal to $S$ complex dimensions or, equivalently, $2S$ real dimensions. Instead of every user sending 4 real, independently encoded streams along 4 linearly independent real signal vectors, suppose users $1, 2, 3$ send $d_1, d_2, d_3$ real, independently encoded streams along $d_1, d_2, d_3$ linearly independent real signal vectors, respectively. As in Section 3, in order to achieve a total of $\frac{d_1+d_2+d_3}{2S}$ degrees of freedom, the received signal vectors for the desired signals must be linearly independent of the received signal vectors carrying interference. The following lemma states a limitation of the alignment scheme.

**Lemma 2** *Suppose vector $\overline{V}_1^1$ aligns with the interference at receivers 2 and 3, i.e.,*

$$\begin{aligned}
\textit{At Receiver 2: } \overline{U}(\phi_{21})\overline{V}_1^1 &= \sum_{s=1}^{d_3} a_s \overline{U}(\phi_{23})\overline{V}_3^s && (69)\\
\textit{At Receiver 3: } \overline{U}(\phi_{31})\overline{V}_1^1 &= \sum_{s=1}^{d_2} b_s \overline{U}(\phi_{32})\overline{V}_2^s && (70)\\
(a_1, a_2, \cdots, a_{d_3}) &\neq (0, 0, \cdots, 0) && (71)\\
(b_1, b_2, \cdots, b_{d_2}) &\neq (0, 0, \cdots, 0) && (72)\\
(a_1, a_2, \cdots, a_{d_3}) &\in \mathbb{R}^{d_3} && (73)\\
(b_1, b_2, \cdots, b_{d_2}) &\in \mathbb{R}^{d_2} && (74)
\end{aligned}$$

*Also, suppose*

$$\sin(\phi_{13} - \phi_{23} + \phi_{21} - \phi_{12} + \phi_{32} - \phi_{31}) \neq 0. \quad (75)$$

*Then $\overline{V}_1^1$ cannot be linearly independent of the interference at receiver 1.*

$$\begin{aligned}
\exists(a'_1, a'_2, \cdots, a'_{d_3}) &\in \mathbb{R}^{d_3} && (76)\\
\textit{and } (b'_1, b'_2, \cdots, b'_{d_2}) &\in \mathbb{R}^{d_2} && (77)\\
\textit{such that} \quad \overline{U}(\phi_{11})\overline{V}_1^1 &= \sum_{s=1}^{d_2} b'_s \overline{U}(\phi_{12})\overline{V}_2^s + \sum_{s=1}^{d_3} a'_s \overline{U}(\phi_{13})\overline{V}_3^s && (78)
\end{aligned}$$

In other words, any given signal vector cannot align with the interference at more than one undesired receivers without becoming aligned within the interference-space at its own desired receiver. Note that if the signal vector becomes aligned within the interference-space at its own desired receiver, then it is useless from a degrees of freedom perspective, i.e., it cannot provide an interference-free signaling dimension.

*Proof:* Given (69) and (70), we wish to show (78). We can express (69) and (70), equivalently, as

$$\begin{aligned}
e^{j\phi_{21}}\overline{\mathbf{V}}_1^1 &= \sum_{s=1}^{d_3} a_s e^{j\phi_{23}}\overline{\mathbf{V}}_3^s && (79)\\
e^{j\phi_{31}}\overline{\mathbf{V}}_1^1 &= \sum_{s=1}^{d_2} b_s e^{j\phi_{32}}\overline{\mathbf{V}}_2^s && (80)
\end{aligned}$$

From Lemma 1 we know that there exist real constants $(c_1, c_2)$ such that

$$\overline{\mathbf{V}}_1^1 = c_1 \mathbf{V}_1^1 e^{j(\phi_{13}-\phi_{23}+\phi_{21}-\phi_{11})} + c_2 \mathbf{V}_1^1 e^{j(\phi_{12}-\phi_{32}+\phi_{31}-\phi_{11})} \quad (81)$$

because the condition

$$\sin(\phi_{13} - \phi_{23} + \phi_{21} - \phi_{12} + \phi_{32} - \phi_{31}) \neq 0 \quad (82)$$

is satisfied by assumption. Substituting from (79), (80) into (81) we have,

$$\overline{\mathbf{V}}_1^1 e^{j\phi_{11}} = c_1 \sum_{s=1}^{d_3} a_s e^{j\phi_{13}} \overline{\mathbf{V}}_3^s + c_2 \sum_{s=1}^{d_2} b_s e^{j\phi_{12}} \overline{\mathbf{V}}_2^s \quad (83)$$

$$\Rightarrow \overline{U}(\phi_{11}) \overline{V}_1^1 = \sum_{s=1}^{d_3} a'_s \overline{U}(\phi_{13}) \overline{V}_3^s + \sum_{s=1}^{d_2} b'_s \overline{U}(\phi_{12}) \overline{V}_2^s \quad (84)$$

$$\text{with } a'_s = c_1 a_s,\ s \in \{1, 2, \cdots, d_3\} \quad (85)$$

$$\text{and } b'_s = c_2 b_s,\ s \in \{1, 2, \cdots, d_2\} \quad (86)$$

which proves the statement of Lemma 2 ■

Lemma 2 highlights a key limitation of the type of linear alignment schemes described in this section. This limitation is formalized in the following theorem.

**Theorem 4** *With the class of linear interference alignment schemes described in this section, the* 3 *user complex Gaussian interference channel with complex channel coefficients, cannot achieve more than* 1.2 *degrees of freedom except over a subset of channel coefficient values of measure* 0*.*

Intuitively, the significance of the number 1.2 can be understood as follows. Consider any signal vector that delivers a coded data stream with one degree of freedom to its desired receiver. It occupies one dimension at its desired receiver. It can share a dimension with an interference vector at one of the receivers where it is undesired, i.e. it can align with interference at one undesired receiver. However, as shown by Lemma 2, it cannot align with interference at the remaining undesired receiver. Thus, it occupies one dimension each at two receivers and half a dimension at the third receiver. The average number of dimensions needed to deliver one degree of freedom is, therefore, $(1+1+0.5)/3 = 2.5/3$. Conversely, the maximum number of degrees-of-freedom delivered per dimension is $3/2.5 = 1.2$. A detailed proof is presented next.

*Proof:* Consider the generalized linear interference alignment scheme, where users $1, 2, 3$ send $d_1, d_2, d_3$ real, independently encoded streams along $d_1, d_2, d_3$ real signal vectors in a $2S$ (real) dimensional vector space created by an $S$ symbol extension of the complex Gaussian interference channel with constant channel coefficients. Because of Lemma 2, we can divide each users' signal space into three disjoint sets. Consider user $i$. The $d_i$ dimensional (real) signal space occupied by transmitter $i$'s signals is represented by the span of the $d_i$ linearly independent columns, $V_i^1, V_i^2, \cdots, V_i^{d_i}$, of the $2S \times d_i$ real matrix $V_i$. This vector space can be divided into three *disjoint* subspaces, spanned by the columns of $\overline{V}_{i1}, \overline{V}_{i2}, \overline{V}_{i3}$ respectively, of size $d_{i1}, d_{i2}, d_{i3}$ such that $d_i = d_{i1} + d_{i2} + d_{i3}$.

$$\text{span}(\overline{V}_i) = \text{span}(\overline{V}_{i1}) \cup \text{span}(\overline{V}_{i2}) \cup \text{span}(\overline{V}_{i3}) \quad (87)$$

$$\text{span}(\overline{V}_{i1}) \cap \text{span}(\overline{V}_{i2}) = \{0\} \quad (88)$$

$$\text{span}(\overline{V}_{i1}) \cap \text{span}(\overline{V}_{i3}) = \{0\} \quad (89)$$

$$\text{span}(\overline{V}_{i2}) \cap \text{span}(\overline{V}_{i3}) = \{0\} \quad (90)$$

$$\text{rank}(\overline{V}_i) = d_i \quad (91)$$

$$\text{rank}(\overline{V}_{ij}) = d_{ij} \quad (92)$$

$$d_i = d_{i1} + d_{i2} + d_{i3} \quad (93)$$

$\forall i, j \in \{1, 2, 3\}$. The partition of the signaling spaces is based on how they align with interference at their unintended receivers. Thus, $\overline{V}_{12}$ is the part of the signal space from transmitter 1 that aligns with the interference from transmitter 2 at receiver 3, $\overline{V}_{13}$ is the signal vector subspace from transmitter 1 that aligns with the interference from transmitter 3 at receiver 2, and the remaining subspace $\overline{V}_{11}$ does not align with interference from any other transmitter at any receiver. The partitioning of signal spaces for transmitter 2 and 3 follows the same interpretation.

$$\text{At Receiver 1: } \text{span}(\overline{U}(\phi_{12})\overline{V}_2) \cap \text{span}(\overline{U}(\phi_{13})\overline{V}_3) = \text{span}(\overline{U}(\phi_{12})\overline{V}_{23} = \text{span}(\overline{U}(\phi_{13})\overline{V}_{32} \quad (94)$$

$$\text{At Receiver 2: } \text{span}(\overline{U}(\phi_{23})\overline{V}_3) \cap \text{span}(\overline{U}(\phi_{21})\overline{V}_1) = \text{span}(\overline{U}(\phi_{23})\overline{V}_{31} = \text{span}(\overline{U}(\phi_{21})\overline{V}_{13} \quad (95)$$

$$\text{At Receiver 3: } \text{span}(\overline{U}(\phi_{31})\overline{V}_1) \cap \text{span}(\overline{U}(\phi_{32})\overline{V}_2) = \text{span}(\overline{U}(\phi_{31})\overline{V}_{12} = \text{span}(\overline{U}(\phi_{32})\overline{V}_{21} \quad (96)$$

Since the signal vectors sent by each transmitter are linear independent among themselves and the channel matrices are invertible, it is easily seen that the following must be true.

$$d_{ij} = d_{ji}, j \neq i, i, j \in \{1, 2, 3\} \quad (97)$$

Note that the partitions of signal spaces outlined above are disjoint. Thus, e.g., there is no subspace of user 1's transmitted signal space that aligns with transmitter 2's interference at receiver 3 *and* also aligns with transmitter 3's interference at receiver 2. This is because Lemma 2 states that such vectors will not be separable from the interference at the desired receiver 1. Since these vectors do not provide interference-free signaling dimensions for user 1, they do not contribute to the degrees of freedom and can be eliminated, as done in the formulation presented above.

Now consider receiver 1. Let us count the total number dimensions spanned by the received signals. The desired signal must be linearly independent of the interference, so it occupies $d_1$ dimensions. The interfering signals from transmitters 2 and 3 have an overlap of $d_{23} = d_{32}$ dimensions, so together they occupy $d_2 + d_3 - d_{23}$ dimensions. Since the total number of dimensions is $2S$, we must have $d_1 + d_2 + d_3 - d_{23} \leq 2S$. Following similar arguments for receivers 2 and 3 we have the following conditions.

$$d_1 + d_2 + d_3 - d_{23} \leq 2S \quad (98)$$

$$d_1 + d_2 + d_3 - d_{31} \leq 2S \quad (99)$$

$$d_1 + d_2 + d_3 - d_{12} \leq 2S \quad (100)$$

Adding these constraints we obtain

$$\frac{d_1 + d_2 + d_3}{2S} - \frac{d_{12} + d_{23} + d_{31}}{6S} \leq 1 \quad (101)$$

Using (97) we bound the second term as follows.

$$d_{12} + d_{23} + d_{31} = \frac{1}{2}(d_{12} + d_{23} + d_{31} + d_{21} + d_{32} + d_{13}) \quad (102)$$

$$\leq \frac{1}{2}(d_1 + d_2 + d_3) \quad (103)$$

Substituting (103) into (101), we obtain

$$\frac{d_1 + d_2 + d_3}{2S}\left(1 - \frac{1}{6}\right) \leq 1 \quad (104)$$

$$\Rightarrow \frac{d_1 + d_2 + d_3}{2S} \leq 1.2 \quad (105)$$

Thus, the total number of degrees of freedom achieved for the $K = 3$ user complex Gaussian interference channel with constant coefficients is no more than 1.2 except over a subset of channel coefficient values of measure 0. ■

# 5 Asymmetric Complex Signaling - Applications

While we introduce the asymmetric complex signaling scheme in this paper with the primary goal of settling the Høst-Madsen-Nosratinia conjecture, the new signaling scheme has broad applications beyond this immediate objective. In this section, we provide a few examples.

## 5.1 Rate Region with Interference as Noise

There is some interest in characterizing the rate region of the interference channel that is achievable by treating interference as noise. For example, [15] characterized the Pareto boundary of the MISO interference channel rate region under this assumption. Using our notation, the basic model for the interference channel used in [15] can be represented as following.

$$\mathbf{Y}_k = \mathbf{H}_{kk}\mathbf{V}_k x_k + \sum_{l=1,l\neq k}^{K} \mathbf{H}_{kl}\mathbf{V}_l x_l + \mathbf{Z}_k \tag{106}$$

where $\mathbf{Y}_k$ is the received complex signal vector, $\mathbf{H}_{kk}$ is the matrix of complex channel coefficients, $\mathbf{V}_k$ is a beamforming vector, $\mathbf{Z}_k$ is the circularly symmetric additive white Gaussian noise vector, and $x_k$ is the *circularly symmetric* complex Gaussian codeword symbol. The achievable rates are then described as:

$$R_k = \log\left(1 + \frac{|\mathbf{H}_{kk}\mathbf{V}_k|^2}{1 + \sum_{l\neq k}|\mathbf{H}_{kl}\mathbf{V}_l|^2}\right) \tag{107}$$

The model described above does not allow the following possibilities.

1. Asymmetric complex signaling
2. Coding over channel extensions
3. Interference Alignment

As shown in this paper, all of these factors have a significant impact on the achievable rates of interference channels, even *with every receiver treating all interference as noise.* Since the single antenna interference channel model studied in this paper can be seen as a special case of the MISO interference channel, and the signaling scheme used in this work also treats interference as noise, it is clear that the rates (107) are suboptimal for the interference channel with single user receivers. In other words, the rate region of interference channels achievable while treating interference as noise is strictly larger than previous characterizations. Interference alignment, channel extensions and most importantly, asymmetric complex signaling will play an important role in solving this problem.

Another related issue is the design of iterative schemes to optimize achievable rates for interference channels, often with the same assumption - treating interference as noise. Even for iterative schemes that do not ignore the possibility of interference alignment, such as the algorithms presented in [16] factoring asymmetric complex signaling into the iterative algorithm may provide higher rates, and as shown in this paper, possibly even higher degrees of freedom.

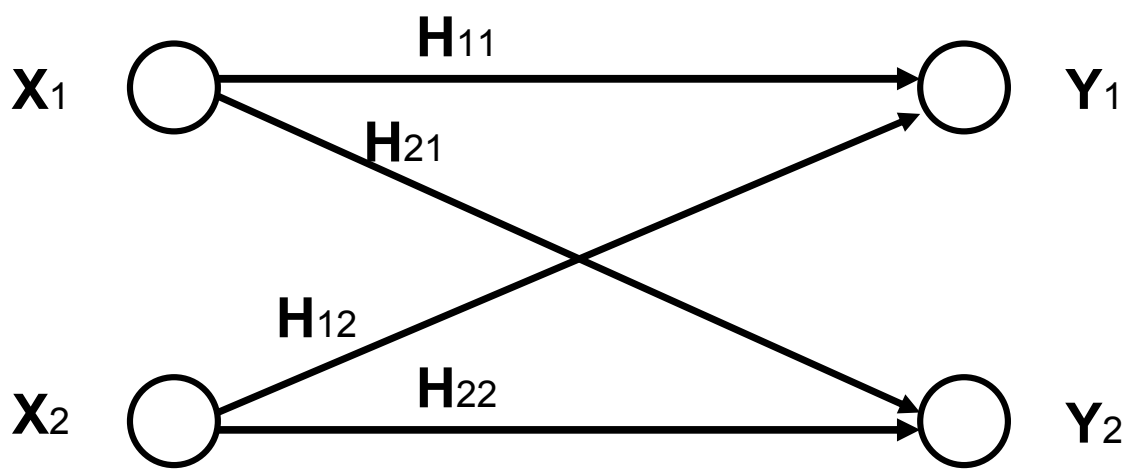

Figure 2: 2-User $X$ Channel

### 5.2 The 2 User $X$ Channel

The 2 user $X$ channel [17] is the same physical channel as the 2 user interference channel. However, in the $X$ channel there are four independent messages, with a message from each transmitter to each receiver. The input-output relationship of the $X$ channel follows from (1) as

$$\mathbf{Y}_r(n) = \mathbf{H}_{r1}\mathbf{X}_1(n) + \mathbf{H}_{r2}\mathbf{X}_2(n) + \mathbf{Z}_r(n) \tag{108}$$

for $r = 1, 2$. Like the interference channel, the $X$ channel can be equivalently represented using real inputs and outputs as

$$Y_r(n) = H_{r1}X_1(n) + H_{r2}X_2(n) + Z_r(n) = h_{r1}U(\phi_{r1})X_1(n) + h_{r2}U(\phi_{r2})X_2(n) + Z_r(n) \tag{109}$$

The message from transmitter $i$ to receiver $j$ is indicated as $W_{ji}$. The rates, capacity and the degrees of freedom of the $X$ channel are defined in a manner similar to the interference channel.

The study of the degrees of freedom of complex Gaussian $X$ channel with constant channel coefficients was pioneered by [4] who showed that if each node is equipped with $M$ antennas then a total of $\lfloor\frac{4M}{3}\rfloor$ degrees of freedom are almost surely achievable. This was a surprising result because the interference channel with the same number of antennas has only $M$ degrees of freedom [18]. The additional degrees of freedom were attributed to an implicit overlap of interference spaces achieved by iterative optimization of transmitters and receivers in [4]. This observation lead to the first explicit interference alignment scheme, introduced in [3]. [3] showed that the constant $X$ channel achieves (almost surely) $\frac{4M}{3}$ degrees of freedom when the nodes have $M$ antennas each. The improvement from $\lfloor\frac{4M}{3}\rfloor$ to $\frac{4M}{3}$ comes not only from the explicit interference alignment scheme but also from a novel idea of channel extensions that was introduced in [13] for the compound broadcast channel and in [3] for the X channel. For the case where each node has only a single antenna, $M = 1$, [3] introduced the idea of channel extensions over time-varying/frequency-selective channels to achieve the outer bound of 4/3 degrees of freedom - this idea was taken further in [2, 6] to establish the degrees of freedom of interference and X networks. However, even with channel extensions the achievability of 4/3 degrees of freedom could not be shown for the *constant* X channel where each node has only a single antenna. The problem, as we show next, was that the achievable scheme was restricted to circularly symmetric signaling. The following theorem shows that with asymmetric complex signaling, the outer bound of 4/3 degrees of freedom is indeed achievable for the 2 user complex Gaussian X channel with constant channel coefficients, for almost all values of channel coefficients.

**Theorem 5** *The $X$ channel has* $4/3$ *degrees of freedom if*

$$\phi_{11} + \phi_{22} \neq \phi_{21} + \phi_{12} \mod (\pi)$$

*Proof:* The converse is proved in [3]. For achievability, we consider a 3 (complex) symbol extension of the channel. This channel may be expressed as

$$\overline{Y}_r(n) = h_{r1}\overline{U}(\phi_{r1})\overline{X}_1(n) + h_{r2}\overline{U}(\phi_{r2})\overline{X}_2(n) + \overline{Z}_r(n)$$

where $\overline{X}_t(n), \overline{Y}_r(n), \overline{Z}_r(n)$ are $6 \times 1$ vectors representing the input, output and additive Gaussian noise respectively over the extended channel. $\overline{U}(\phi_{rt})$ represents the block-diagonal channel matrix determined by $\phi_{rt}$ - the phase of the channel gain between transmitter $t$ and receiver $r$. Over this extended channel, 2 interference free streams are achieved for each of the 4 messages using beamforming. Now, let $V_{ij}$ be the $6 \times 2$ matrix whose columns are used by transmitter $j$ as beamforming directions for message $W_{ij}$. The achievable scheme mimcs the scheme provided for time-varying channels in [3], i.e., the vectors are chosen so that, all vectors meant for receiver 1 align at receiver 2 and *vice-versa*. Specifically, matrices $\overline{V}_{11}, \overline{V}_{21}$ are chosen randomly from any continuous distribution. Then $\overline{V}_{12}, \overline{V}_{22}$ are chosen to satisfy the following alignment conditions

$$\overline{U}(\phi_{22})\overline{V}_{12} = \overline{U}(\phi_{21})\overline{V}_{11} \quad (110)$$

$$\overline{U}(\phi_{12})\overline{V}_{22} = \overline{U}(\phi_{11})\overline{V}_{21} \quad (111)$$

The above equations ensure that the 4 interfering vectors at receiver $i \in \{1, 2\}$ align into the 2 dimensional space represented by $\overline{U}(\phi_{i1})\overline{V}_{j1}$ where $j \neq i$. Thus a receiver can resolve the 4 dimensions corresponding to the desired streams, provided that the desired signal space is linearly independent of the interference signal space. Therefore, at receiver 1, we need to ensure that the following 6 vectors are linearly independent.

$$\overline{U}(\phi_{11})\overline{V}_{11}^1, \overline{U}(\phi_{11})\overline{V}_{11}^2, \overline{U}(\phi_{12})\overline{V}_{12}^1, \overline{U}(\phi_{12})\overline{V}_{12}^2, \overline{U}(\phi_{11})\overline{V}_{21}^1, \overline{U}(\phi_{11})\overline{V}_{21}^2 \quad (112)$$

where $\overline{V}_{ij}^1$ and $\overline{V}_{ij}^2$ are $6 \times 1$ column vectors representing the two columns of $\overline{V}_{ij}$. To show that the above set of equations are linearly independent, assume the contrary, i.e., assume that there exist real constants $a_1, a_2, \ldots, a_6$ not all equal to zero, so that

$$a_1\overline{U}(\phi_{11})\overline{V}_{11}^1 + a_2\overline{U}(\phi_{11})\overline{V}_{11}^2 + a_3\overline{U}(\phi_{12})\overline{V}_{12}^1 + a_4\overline{U}(\phi_{12})\overline{V}_{12}^2 + a_5\overline{U}(\phi_{11})\overline{V}_{21}^1 + a_6\overline{U}(\phi_{11})\overline{V}_{21}^2 = 0$$

Now, using (111) above and simplifying, we get

$$a_1\overline{V}_{11}^1 + a_2\overline{V}_{11}^2 + a_3\overline{U}(\phi_{12} + \phi_{21} - \phi_{22} - \phi_{11})\overline{V}_{11}^1 + a_4\overline{U}(\phi_{12} + \phi_{21} - \phi_{22} - \phi_{11})\overline{V}_{11}^2 + a_5\overline{V}_{21}^1 + a_6\overline{V}_{21}^2$$

Since $\phi_{11} + \phi_{22} \neq \phi_{12} + \phi_{21} \mod (\pi)$, noting that $a_i, i = 1, 2, \ldots, 6$ are real and taking the imaginary part of the above equation, we get

$$a_3 \sin(\phi_{12} + \phi_{21} - \phi_{22} - \phi_{11})\overline{V}_{11}^1 + a_4 \sin(\phi_{12} + \phi_{21} - \phi_{11} - \phi_{22})\overline{V}_{11}^2 = 0$$

Since $V_{11}^1, V_{11}^2$ are $6 \times 1$ vectors chosen randomly, they are linearly independent almost surely. Therefore, we get $a_3 = a_4 = 0$. Using this in (112), we get

$$a_1\overline{V}_{11}^1 + a_2\overline{V}_{11}^2 + a_5\overline{V}_{21}^1 + a_6\overline{V}_{21}^2 = 0$$

Again, note that $\overline{V}_{11}^1, \overline{V}_{11}^2, \overline{V}_{21}^1, \overline{V}_{21}^2$ are four $6 \times 1$ vectors, all chosen independent of each other from continuous probability distributions, and must therefore be linearly independent almost surely. Thus we have $a_i = 0, i = 1, 2, \ldots, 6$. This implies that the desired signal dimensions are linearly independent of the interfering dimensions almost surely at receiver 1. Further, by symmetry of construction, we can claim that the 4 desired signal streams are linearly independent of the 2 interfering directions at receiver 2 as well. This ensures that 8 interference free streams are achievable over 6 real dimensions over the extended $X$ channel. Thus the number of degrees of freedom achieved per channel use is $4/3$.

### 5.3 Cognitive X Channel

Without loss of generality, an X channel with a cognitive receiver (transmitter) is one where, e.g., a genie provides receiver (transmitter) 2 with the message $W_{11}$. It is shown in [3] that for the complex Gaussian $X$ channel with a cognitive receiver (transmitter) and constant channel coefficients, if each node has $M > 1$ antennas, then the number of degrees of freedom is 3/2. The question is left open for $M = 1$ in [3] if the channel coefficients are constant. However, it is easily seen that using asymmetric complex signaling, 3/2 degrees of freedom are achieved even for $M = 1$, for both cognitive transmitter, and cognitive receiver X channel models. The achievable scheme is essentially identical to the one proposed in [3] for $M > 1$, except the multiple signaling dimensions come not from multiple antennas but are inherent in the complex symbols, and the alignment happens in channel phase. We summarize the asymmetric complex signaling based alignment scheme for the cognitive receiver case as follows. Set the rate for message $W_{12}$ to zero, and view a complex symbol as a two dimensional vector space of real vectors. Now, transmitters 1 and 2 send messages $W_{21}, W_{22}$ encoded with real Gaussian codebooks, and phase rotated so that they are aligned at receiver 1. Because of the relativity of alignment, these vectors are almost surely separable at receiver 2. Transmitter 1 also separately encodes and sends $W_{11}$ so that it is received orthogonal to the aligned interference vectors at receiver 1. Receiver 2 is able to eliminate the interference from the codeword for $W_{11}$ because it knows $W_{11}$. Thus, a total of 3 real signaling dimensions are created over one complex symbol, i.e. over two real symbols, i.e., 3/2 degrees of freedom are achieved. The cognitive transmitter case follows similarly by a combination of asymmetric complex signaling and the achievable scheme proposed in [3].

### 5.4 Cellular Application - Interfering Uplinks

Consider two interfering 2-user multiple access channels (See Fig. 3). This channel can be used to model two interfering cells in a cellular network [19]. The channel has 4 transmitters and 2 receivers with input-output relations as below.

$$Y_1(n) = \sum_{k=1}^{4} h_{1k} U(\phi_{1k}) X_k(n) + Z_1(n) \tag{113}$$

$$Y_2(n) = \sum_{k=1}^{4} h_{2k} U(\phi_{2k}) X_k(n) + Z_2(n) \tag{114}$$

(115)

Transmitters $1, 2$ each have a message for receiver 1 and transmitters $3, 4$ have a message for receiver 2. Let $W_i$ represent the message present at tramsnmitter $i$. The power constraint, rate-region, degrees of freedom of this channel are in the same manner as the interference channel. We indicate the rate and the degrees of freedom of message $W_i$ by $R_i$ and $d_i$ respectively, for $i = 1, 2, 3, 4$. We characterize the degrees of freedom of the interfering multiple access channels below.

**Theorem 6** *If*

$$\begin{aligned} \phi_{11} + \phi_{22} &\neq \phi_{21} + \phi_{12} \mod (\pi) \\ \phi_{23} + \phi_{14} &\neq \phi_{13} + \phi_{24} \mod (\pi) \end{aligned}$$

*Then the interfering uplinks model described above has* $\frac{4}{3}$ *degrees of freedom.*

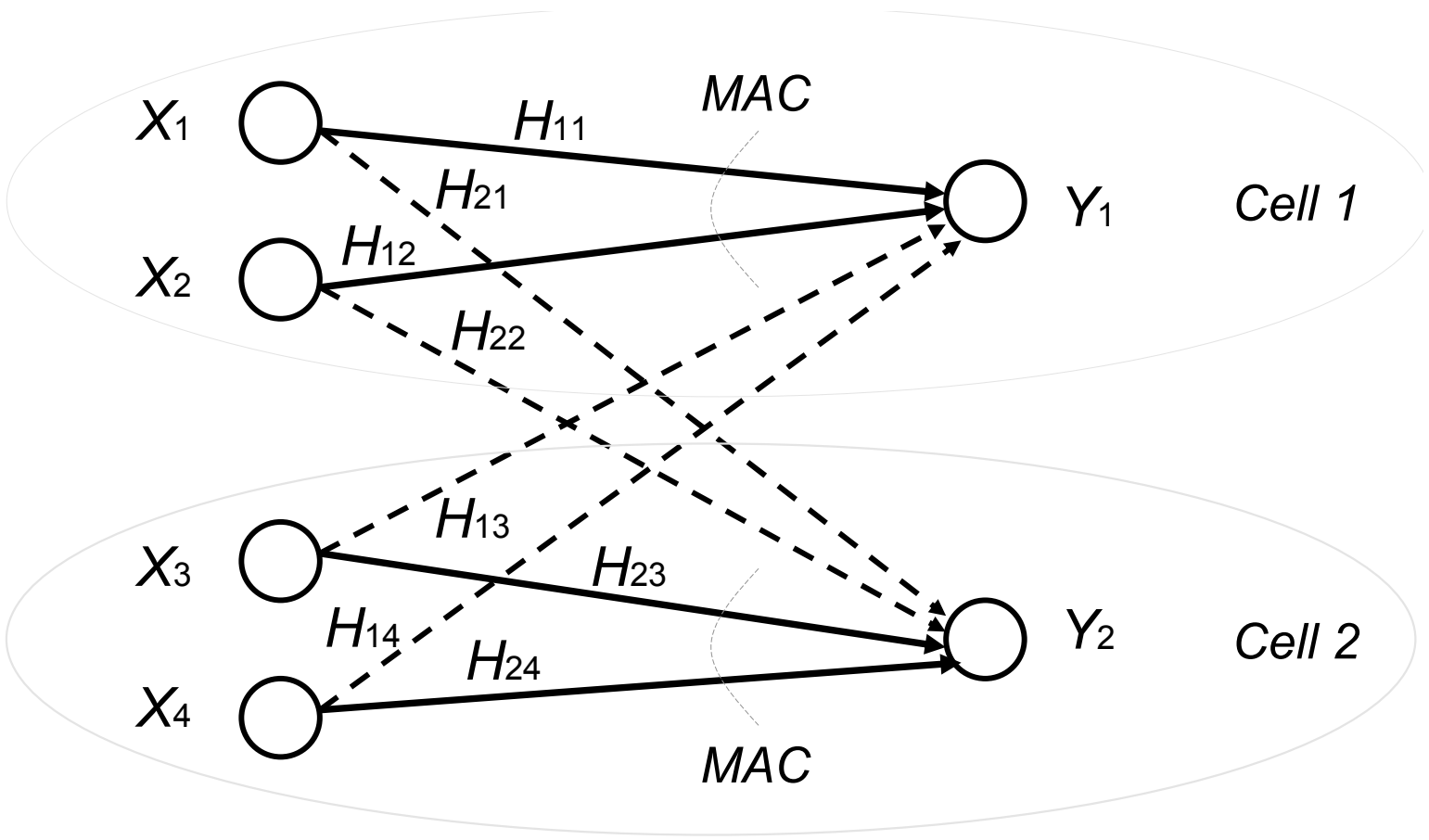

Figure 3: Two Interfering 2-User Multiple Access Channels

#### 5.4.1 Proof of Converse:

We show the following relations

$$\begin{aligned} d_1 + d_3 + d_4 &\leq 1 \\ d_2 + d_3 + d_4 &\leq 1 \\ d_1 + d_2 + d_3 &\leq 1 \\ d_1 + d_2 + d_4 &\leq 1 \end{aligned}$$

Summing all the relations above, we can bound the total number of degrees of freedom of the channel by 4/3. We only show the first inequality here. The remaining 3 inequalities follow from symmetry. Now, to show the inequality, set $W_2 = \phi$. Note that setting certain messages to null does not decrease the degrees of freedom achieved by other messages [3]. Now, let a genie provide $W_3, W_4$ to receiver 1. Receiver 1 can now cancel the signals from transmitters $3, 4$ to obtain

$$Y_1^{'}(n) = h_{11}U(\phi_{11})X_1(n) + Z_1(n)$$

Note that receiver 1 can decode $W_1$ using $Y_1^{'}$. Now, using any achievable scheme, receiver 2 can decode $W_3, W_4$ and therefore cancel the effect of $X_3, X_4$ to obtain

$$Y_2^{'}(n) = h_{21}U(\phi_{21})X_1(n) + Z_2(n)$$

Since all noise variables are circularly symmetric, by reducing the noise variance of $Z_2$ sufficiently, we can ensure that $Y_1^{'}$ is a degraded version of $Y_2^{'}$. Note that reducing the noise variance does not reduce the degrees of freedom region. Receiver 2 can now decode $W_1$ as well, so that the rates of the messages $W_1, W_3, W_4$ lie in a multiple access channel formed at an enhanced receiver 2. Since the multiple access channel has 1 degree of freedom, we can write

$$d_1 + d_3 + d_4 \leq 1$$

This completes the proof of the converse ■

### 5.4.2 Proof of Achievability:

Consider a 3 symbol extension of the channel. This channel may be expressed as

$$\begin{aligned}\overline{Y}_1 &= \sum_{j=1}^{4} h_{1j}\overline{U}(\phi_{1j})\overline{X}_j + \overline{Z}_1 \\ \overline{Y}_2 &= \sum_{j=1}^{4} h_{2j}\overline{U}(\phi_{2j})\overline{X}_j + \overline{Z}_2\end{aligned}$$

where $\overline{X}_t, \overline{Y}_r, \overline{Z}_r$ are $6 \times 1$ vectors representing the input, output and additive Gaussian noise respectively over the extended channel. $\overline{U}(\phi_{rt})$ represents the block-diagonal channel matrix determined by $\phi_{rt}$ - the phase of the channel gain between transmitter $t$ and receiver $r$. Over this extended channel, 2 interference free streams are achieved for each of the 4 messages using beamforming. Let $V_j$ be the $6 \times 2$ matrix whose columns are used by transmitter $j$ as beamforming directions for message $W_j$. Like the $X$ channel vectors are chosen so that, all vectors meant for receiver 1 align at receiver 2. Specifically, let $\overline{V}_1, \overline{V}_3$ be two $6 \times 2$ real matrices chosen randomly from any continuous distribution. Then $\overline{V}_i$ are chosen to satisfy the following alignment conditions

$$\overline{U}(\phi_{21})\overline{V}_1 = \overline{U}(\phi_{22})\overline{V}_2 \tag{116}$$

$$\overline{U}(\phi_{13})\overline{V}_3 = \overline{U}(\phi_{14})\overline{V}_4 \tag{117}$$

Note that the above equations ensure that all 4 vectors at receiver $i \in \{1, 2\}$, align in a 2 dimensional space. Now, all we need to ensure is that at receiver $i$, the desired signalling directions are linearly independent of the interfering directions. Now, consider the vectors received at receiver 1.

$$\overline{U}(\phi_{13})\overline{V}_3, \overline{U}(\phi_{11})\overline{V}_1, \overline{U}(\phi_{12})\overline{V}_2$$

Note that the above vectors can be equivalently represented as

$$\overline{U}(\phi_{13})\overline{V}_3, \overline{U}(\phi_{11})\overline{V}_1, \overline{U}(\phi_{12} + \phi_{21} - \phi_{22})\overline{V}_1$$

We need to show that all the 6 vectors are linearly independent of each other. The proof is similar to the $X$ channel. In other words, consider real constants $a_i, i = 1, 2, \ldots 6$ such that

$$a_1\overline{U}(\phi_{13})\overline{V}_3^1 + a_2\overline{U}(\phi_{13})\overline{V}_3^2 + a_3\overline{U}(\phi_{11})\overline{V}_1^1 + a_4\overline{U}(\phi_{11})\overline{V}_1^2 + a_5\overline{U}(\phi_{12}+\phi_{21}-\phi_{22})\overline{V}_1^1 + a_6\overline{U}(\phi_{12}+\phi_{21}-\phi_{22})\overline{V}_1^2 = 0 \tag{118}$$

where $\overline{V}_i^1, \overline{V}_i^2$ are the two column vectors of $\overline{V}_i$. Now, the argument that $a_i = 0, i = 1, 2, 3 \ldots 6$ is almost identical to the argument presented for the $X$ channel and we will only present the outline here for brevity. Now, we consider 2 cases.

*Case 1*: $\sin(\phi_{13} - \phi_{11}) \neq 0$

Then, mutliplying (118) by $\overline{U}(-\phi_{11})$ and equating the imaginary part to 0, we get 6 equations in $a_1, a_2, a_5, a_6$ since $\sin(\phi_{11} + \phi_{22} - \phi_{21} - \phi_{12}) \neq 0$. Since vectors are chosen randomly, we can show that $a_1 = a_2 = a_4 = a_6 = 0$. Then, using this in (118) again, we can show that $a_1 = a_2 = 0$.

*Case 2*: $\sin(\phi_{13} - \phi_{11}) = 0$

Note that this means that $\cos(\phi_{13} - \phi_{11}) \neq 0$. Then, equating the imaginary part of the left-hand-side of (118) to 0, we can show that $a_5 = a_6 = 0$. Plugging this back into (118) and taking its real part, we can show that $a_1 = a_2 = a_3 = a_4 = 0$. This shows that the desired signals are linearly independent of the interference at receiver 1. By symmetry of construction, we can show that the desired vectors are linearly independent of the interference at receiver 2 as well. ■

## 6 Conclusion

We settle the Høst-Madsen-Nosratinia conjecture in the negative, by establishing that complex Gaussian interference networks with more than 2 users and constant channel coefficient coefficients have at least 1.2 degrees of freedom for almost all values of channel coefficients. The achievability of 1.2 degrees of freedom is based on interference alignment with only 3 simultaneously active users employing asymmetric complex signaling over supersymbols consisting of 5 complex channel symbols. The main limitation of this alignment scheme, for the 3-user case, is that each signal vector can only align with interference at no more than one undesired receiver, which translates into the maximum of 1.2 degrees of freedom for the 3 user interference channel. The scheme is shown to achieve the degrees-of-freedom outer bound for the 2 user complex Gaussian $X$ channel with constant coefficients, thereby improving upon previous results which relied on time-varying/frequency-selective channel coefficients. An interesting feature of this alignment scheme is that it is concerned only with the phase and not at all with the magnitudes of the channel coefficients. Remarkably, this is the opposite of all previously considered signal *level* based alignment schemes for constant channels, which are concerned primarily with the magnitudes of the channel coefficients and are essentially restricted to real channel coefficients.

The degrees-of-freedom of real Gaussian interference channels with constant channel coefficients remain open for almost all channel coefficient values. However, because wireless channels are invariably modeled as complex Gaussian, the more interesting question may be to determine if more than 1.2 degrees of freedom can be achieved for non-negligible subset of complex Gaussian constant channels. For $K = 3$ users, it may require smart ways of combining signal level alignment schemes (that exploit the variety of channel magnitudes) and signal vector space alignment schemes (that exploit the variety of channel phases) . For more than 3 users, it will be interesting to determine the limitations of interference alignment with asymmetric complex signaling over constant channels. Using asymmetric complex signaling to improve existing interference alignment schemes [2] as well as to design more efficient iterative algorithms [16] are also promising directions for future work.

Beyond the degrees of freedom problem, the key new insight to emerge from this work is the idea of asymmetric complex signaling. We expect this fundamental idea may have a variety of applications and point out some examples in this paper. In conclusion, along with interference alignment [2], need for channel extensions [3], and inseparability of parallel interference channels [12], the need for asymmetric complex signaling can be added as yet another essential piece of the puzzle that is the capacity of interference networks.

## Acknowledgment

The authors would like to acknowledge Tiangao Gou, whose comments helped simplify the achievable scheme presented in this paper.